# Anisotropic transverse magnetoresistance temperature dependence in Mn$_3$Ga Weyl semimetal with chiral anomaly

Isis M. Cota-Martínez,[†] Ricardo López Antón,[§] Andrés M. Garay-Tapia,[†] José A. Matutes-Aquino,[†] Carlos R. Santillán-Rodríguez,[†] Renee J. Saénz-Hernández,[†] Rocío M. Gutiérrez-Pérez,[♯] José T. Holguín-Momaca,[†] Caroline A. Ross,[‡] Sion F. Olive-Méndez[†]*

[†]Centro de Investigación en Materiales Avanzados, S.C. (CIMAV), Miguel de Cervantes No. 120, C.P. 31136 Chihuahua, Mexico

[‡]Department of Materials Science and Engineering, Massachusetts Institute of Technology, 77 Massachusetts Avenue, Cambridge, Massachusetts 02139, USA

[§]Instituto Regional de Investigación Científica Aplicada (IRICA) and Departamento de Física Aplicada, Universidad de Castilla-La Mancha, 13071 Ciudad Real, Spain

[♯]Facultad de Ciencias Químicas, Universidad Autónoma de Chihuahua, Circuito Universitario s/n Campus II, Chihuahua, Mexico

**ABSTRACT**: Hexagonal antiferromagnetic D0$_{19}$-Mn$_3$*X* (*X* = Sn, Ge, Ga) compounds, with a non-collinear Kagome spin structure, are Weyl semimetals exhibiting novel topological transport properties. The longitudinal magnetoresistance of *c*-oriented epitaxial Ru/Mn$_3$Ga thin films exhibits a positive quadratic dependence on magnetic field over a wide range of temperatures. Here we describe the transverse magnetoresistance, with the field in the out-of-plane direction, for *c*-oriented epitaxial GaN (0001)/Mn$_3$Ga films. There is a transition from a negative linear to a positive quadratic dependence on magnetic field in the temperature range from 200 K to 300 K. The electrical resistivity shows a metallic to semiconductor transition at 230 K. By applying the electric field along two perpendicular in-plane directions we find asymmetry in the magnetoresistance curves due to self-spin polarized currents created through magnetic octupole clusters. First principles calculations confirmed the metallic to semiconductor transition corresponds to reordering the spin structure to a higher symmetry configuration.

Keywords: Magnetoresistance, Weyl semimetals, Chiral anomaly, Kagome lattice, metal, semiconductor.

Antiferromagnetic materials have recently attracted worldwide attention due to their advantages over conventional ferromagnets such as their magnetic stability upon temperature, magnetic and electric field perturbations. Furthermore, they offer faster spin dynamics, in the THz regime, and the lack of stray fields, which make them suitable for the design of robust and ultrafast spintronic devices.[1,2,3] More recently, antiferromagnets with non-collinear spin structures as the MnSi and MnGe helimagnets have revealed skyrmion lattices and the observation of topological Hall effect.[4,5,6,7,8] Another family of materials with topological properties consists of the $D0_{19}$-Mn$_3$X (X = Sn, Ge, Ga) hexagonal isostructural compounds (space group P6$_3$/mmc). They have been subjects of intense research after the observation of anomalous and topological Hall effects in bulk single crystals and polycrystalline ingots.[9,10,11] In particular, Mn$_3$Ga possesses a Kagome spin structure, with the Ga atom located at the center of the hexagon, under the magnetic space group 63.464, which is the most accepted theoretically calculated spin structure.[12,13] In the case of the Mn$_3$Ge, the anomalous Hall effect is attributed to the non-vanishing Berry curvature in momentum space.[11] Meanwhile, the topological Hall effect in bulk Mn$_3$Ga is attributed to a change of the spin texture induced by a hexagonal to orthorhombic distortion at low temperatures.[14] The topological Hall effect is also attributed to the non-zero scalar spin chirality $\Omega = s_1 \cdot (s_2 \times s_3)$, where $s_i$ are the magnetic moments of adjacent atoms.[15] This chirality is non-zero in Mn$_3$Ga, in the absence of a magnetic field, as the Mn moments lying in the (0001) basal plane exhibit a canting of ~2-3° along the $c$ axis producing a ferromagnetic component.[16] Transverse and longitudinal magnetoresistance (MR) has been measured in Mn$_3$Ga epitaxial thin films and bulk single crystals, being either positive or negative and described by a quadratic or linear dependence on the magnetic field, $B$, depending on the crystal orientation of the Mn$_3$Ga and the orientations of the magnetic and electric fields.[17,18,19,20] Recently, all-antiferromagnetic magnetic tunnel junctions based on Mn$_3$Sn/MgO revealed a ~2% room-temperature tunneling magnetoresistance driven by magnetic octupoles.[21] Magnetic octupoles in D0$_{19}$ antiferromagnets have a projected-density of states with strong polarization of opposite octupole moments in the Kagome plane. Octupole clusters have acted as self-spin polarizers in polycrystalline Mn$_3$Sn thin films.[22,23,24] Finally, the Néel temperature of bulk Mn$_3$Ga is 470±10 K, the highest among the D0$_{19}$-Mn$_3$X compounds, making it a suitable choice for room-temperature spintronic applications.[11,25,26,27]

Here we report on the structural, magnetic and magnetotransport properties of *c*-oriented epitaxial D0$_{19}$-Mn$_{2.46}$Ga thin films, focusing on the sign of the magnetoresistance together with its linear or quadratic dependence on the magnetic field. The electric field induced electric current, *I*, is directed in two perpendicular directions within the basal Mn$_3$Ga (0001) plane with the magnetic field perpendicular to the film plane. We explore the magnetic properties of Mn$_3$Ga and discuss the magnetoresistance curves in terms of the interaction of *I* with the Kagome lattice and the effect of the magnetic octupoles as self-spin polarizers. First principles calculations agree well with the experimental observation of a metallic to semiconductor transition by proposing an evolution of the spin structure, from the semimetallic magnetic space group 63.464 to a higher symmetry state described by the magnetic space group 176.147.

The GaN (0001)/Mn$_3$Ga epitaxial thin films, with thickness of 70 nm, were grown by rf-magnetron sputtering from a stoichiometric Mn$_3$Ga target at a substrate temperature of 400 °C. The single crystal quality of the films was confirmed *in-situ* with reflection high-energy electron diffraction (RHEED) and the phase identification was performed by X-ray diffraction (XRD). The composition of the film was determined by energy dispersive X-ray spectroscopy (EDS). The magnetization-magnetic field (*M-H*) hysteresis loops were acquired using a superconducting quantum interference device (SQUID) magnetometer and the magnetoresistance (MR), resistivity and temperature-dependent magnetization were measured in a physical property measuring system (PPMS). In all the measurements, the applied magnetic field was oriented out-of-plane. The films were patterned into 100 μm wide bars using direct-write laser lithography with positive resist and plasma etching. Transverse MR was measured with $I = 7.5$ mA (~$10^9$ Am$^{-2}$) under an applied magnetic field of 5 T. The calculations were performed using density functional theory (DFT)[28,29] as implemented in the Vienna ab-initio simulation package (VASP).[30,31] Wave functions were expanded in plane waves, using only valence electrons described within the projector augmented-wave (PAW)[32] pseudopotential formalism using the SCAN[33] metaGGA functional. A gamma-centered k point mesh of 10×10×10 and a kinetic energy cut-off of 500 eV for the plane wave basis set were used for full relaxation. These values ensured the convergence of residual forces of 0.001 eV·A. For self-consistent calculations, a k-mesh of 16×16×16 was used and the electronic minimization was set to 1×10$^{-7}$. In order to consider non-collinear magnetic configurations, spin-orbit coupling (SOC) was included in all calculations.[34]

The D0$_{19}$-Mn$_3$Ga crystal structure along with its non-collinear spin structure is shown in Fig. 1(a). The lattice constants from first principles calculations, performed in this work for fully stoichiometric bulk Mn$_3$Ga, are $a = 5.42$ Å and $c = 4.32$ Å, which are close to results from other studies.[13,25] Our Mn$_3$Ga thin films were epitaxially grown on GaN (0001) templates on *c*-cut sapphire substrates. The XRD pattern acquired in Bragg-Brentano configuration shown in Fig. 1(b) exhibits peaks of GaN (0001), *c*-cut sapphire and the (0002) and the (0004) peaks of Mn$_3$Ga located at 40.66° and 89.98°, which implies $c = 4.394 \pm 0.054$ Å. The rocking curves of the GaN (0002) and Mn$_3$Ga (0002) peaks of a bare substrate and a Mn$_3$Ga thin film shown in Fig. 1(c) were collected to identify the crystalline quality. The full width at half maximum of those peaks are 0.053° and 1.548°, respectively, indicating the high crystalline quality of the GaN template layer and the presence of dislocations or other defects in the film to accommodate the lattice mismatch with the substrate. The RHEED patterns, shown in Figs. 1(d) and 1(e), consist of sharp spots along the $[11\bar{2}0]$ and $[\bar{1}100]$ directions, respectively. The arrangement of spots in the RHEED patterns represents the surface of a single crystalline thin film with hexagonal crystal structure, *i.e.* the (0001) surface of Mn$_3$Ga.[35,36] The spot-like RHEED patterns indicates that electron diffraction takes place in transmission through a 3-dimensional surface morphology. Atomic force microscopy revealed that epitaxial growth is achieved within an island-like growth topography with a rms surface roughness of 2.9 nm as shown in Fig. 1(f). RHEED patterns of substrate and film were correlated to identify the hexagon-on-hexagon epitaxy in which a $(\sqrt{3} \times \sqrt{3})R30°$ GaN surface unit cell ($\sqrt{3}a_{\text{GaN}} = 5.52$ Å) matches a unit cell of Mn$_3$Ga, yielding the epitaxial relationship GaN $(0001)[1\bar{1}00]\|$ Mn$_3$Ga $(0001)[11\bar{2}0]$. The chemical composition of the film determined by EDS is Mn$_{2.88\pm0.14}$Ga (denoted hereinafter as Mn$_3$Ga).

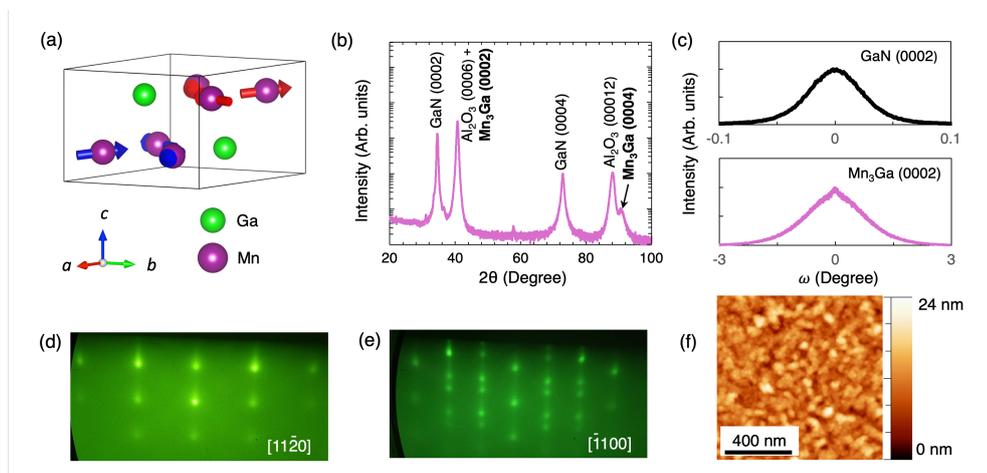

**Figure 1.** (a) VESTA visualization of the crystal structure of D0$_{19}$-Mn$_3$Ga.[37] (b) XRD pattern of the *c*-cut sapphire/GaN (0001)/Mn$_3$Ga (0001) heterostructure. RHEED patterns of the 75-nm thick Mn$_3$Ga (0001) surface along the (c) $[11\bar{2}0]$ and (b) $[\bar{1}100]$ directions of Mn$_3$Ga.

The in-plane (IP) *M-H* loops at 300 K measured along three azimuthal angles of 0°, 45° and 90° with respect to the IP [11$\bar{2}$0] direction, are shown in Fig. 2(a). The loops overlap with equal shape, presenting zero coercivity $H_c$ and saturation magnetization, $M_s$ of 164 kAm$^{-1}$. The zero coercivity agrees well with the evidence that the Kagome lattices of Mn$_3$Sn and Mn$_3$Ge rotate in the film plane with the direction of the magnetic field.[38] To further explore the dependence of $M_s$ and coercivity $H_c$ on the field, (out-of-plane) OP *M-H* loops at 150, 200, 250, and 300 K, with a maximum applied field of 5 T, were collected.

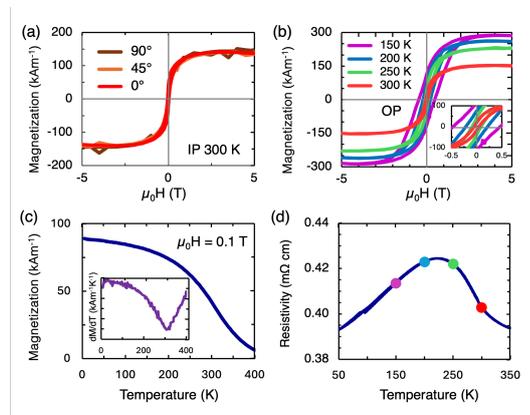

**Figure 2.** Magnetization vs. applied field (a) in the film plane along three azimuths from the direction [11$\bar{2}$0] at 300 K. (b) Out-of-plane hysteresis loop at different temperatures. (c) Magnetization vs. temperature at 0.1 T applied field. The inset represents the derivative of the *M-T* curve. (d) Resistivity vs. temperature curve with a metallic to semiconductor transition at 230 K. The dots represent the temperatures at which the magnetoresistance was measured.

The results in Fig. 2(b) depict an increase in both $M_s$ and $H_c$ as the temperature decreases, a signature of the OP ferromagnetic component of Mn$_3$Ga. In particular, $M_s$ increases from 150 to 280 kAm$^{-1}$ at temperatures of 300 and 150 K, respectively, corresponding to a magnetic moment of 0.24 $\mu_B$/Mn at 300 K. This moment is larger than that of thinner Mn$_3$Ga films (of 10 and 15 nm) grown on a Ru buffer layer, producing exchange bias with CoFe ultrathin films, of 0.09 $\mu_B$/Mn due to a low canting of 2-3° of the magnetic moments[16] and 40 nm-thick *ab*-axis oriented thin films grown on W (211) buffer layer have a magnetic moment of 1.5 $\mu_B$/Mn at 300 K.[18] The lower lattice mismatch of Mn$_3$Ga with the Ru buffer, $2a_{Ru}$ = 5.412 Å, $\Delta a/a$ = -0.1% reduces dislocation density, which in turn favors the growth of films with high crystalline quality thus preserving the long range Kagome ordering.[39] In our GaN (0001)/ Mn$_3$Ga (0001) system, the Mn vacancies and dislocations may contribute to break the long range Kagome antiferromagnetic ordering, thus increasing the net magnetic moment per Mn. Figure 2(c) shows the magnetization vs. temperature, *M-T*, curve from 5 to 400 K, measured under an applied magnetic field of 0.1 T, where the magnetization decreases monotonically with increasing temperature.

The shape of the curve shows the same trend as that of bulk Mn$_3$Ga, which is attributed to the canting of the Mn moments.[40] The strength of the magnetic interactions starts dimishing after 309 K as shown in the inset, however, short range interactions preserve the OP magnetization of the Kagome lattice beyond 400 K. Therefore, the Néel temperature is still far above the range at which MR measurements were carried out, as will be shown later (bulk $T_N$ = 470±10 K).[25]

Mn$_3$Ga undergoes a structural distortion from hexagonal to orthorhombic on cooling the sample from room temperature, at 120 K in polycrystalline Mn$_{2.44}$Ga ribbons and at 170 K in Mn$_{2.85}$Ga$_{1.15}$ powders.[20,41] In both systems, the transition temperature is manifested as a strong variation of *M* in the *M-T* curves. However, in our samples this transition is not observed in the *M-T* curve (see Fig. 2 (c)) as the film is constrained by epitaxy with the substrate thus preserving the hexagonal structure even at temperatures below 120 K.

Figure 2(d) shows the resistivity *vs*. temperature, $\rho$-*T*, curve in the range from 50 to 350 K; the dots along the $\rho$-*T* curve are the temperatures at which the MR curves were measured. The shape of the curve indicates a transition from metallic to semiconductor behavior as the temperature increases, with a maximum resistivity at 230 K. This temperature is associated with a transition in the spin structure from the *Cm'cm'* (63.464) to the *P6₃'/m'* (176.147) magnetic space group, which is the spin structure with the highest symmetry.[42,43] A semiconductor to metallic transition has been observed in $Mn_3Sn$ through a commensurate-incommensurate magnetic phase transition.[44] However, for our $Mn_3Ga$ films, the metallic to semiconductor transition is associated with a incommensurate-commensurate transition in the spin structure.[44] At elevated temperatures, the competition between different spin domains modifies the magnetic and electronic properties as discussed below.

The top-view of $D0_{19}$-$Mn_3Ga$ (see Fig. 3(a)) shows the Kagome lattices in $z = ¼$ and $¾$ (red and blue spin sublattices) along with the unit cell and the octupole. The pink and blue arrows are the $[11\bar{2}0]$ and the $[\bar{1}100]$ directions in which current *I* flowed to measure the transverse MR, $\Delta\rho(H) = [(\rho(B) - \rho(B = 0))/\rho(B = 0)] \times 100$. In the following, the transverse MR will be denoted as MR1 (measured in the $[11\bar{2}0]$ direction) and MR2 (measured along $[\bar{1}100]$). In these measurements, the magnetic field remains perpendicular to the film plane. RHEED was used to confirm the orientation of the substrate and identify the MR1 or MR2 orientation of the patterned bars used to measure MR.

The results of MR2 devices at 150, 200, 250, and 300 K are shown in Fig. 3(b). The MR curves measured at 150 and 200 K, fully in the metallic regime of the sample, are negative; a higher negative MR ratio can be obtained by reducing the temperature from 200 to 150 K. In this temperature range, the MR dependence on the field is linear, which corresponds to the behavior of a Weyl semimetal type II subject to a chiral anomaly.[10,45,46] This chiral anomaly consists on the tilting of the of the Weyl cones at the nodes where $e^-$ and $h^+$ densities intersect. The positive magnetoconductance, $\Delta\sigma(H)$, is related to the negative magnetoresistance through $\Delta\sigma(H) = -\Delta\rho(H)/\rho_{xx}^2$. The chiral anomaly represented by positive magnetoconductance is the signature of magnetic Weyl fermions. The prediction of Weyl type-II fermions from $Mn_3Ge$ and $Mn_3Sn$ was confirmed by experimental evidence of longitudinal linear positive magnetoconductance and negative transverse magnetoconductance due to the chiral anomaly.[11,47]

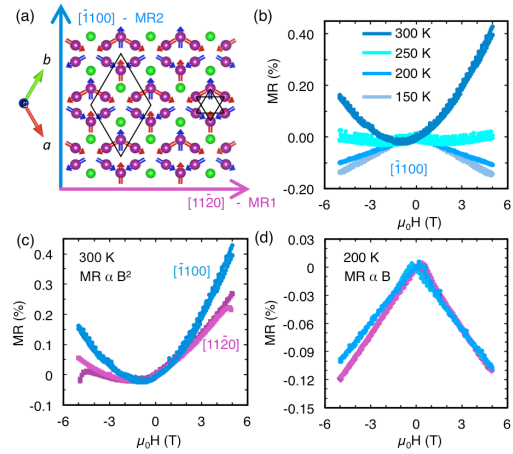

**Figure 3.** (a) Kagome lattices on the (0001) plane of Mn$_3$Ga with the directions MR1 and MR2 on which the transverse magnetoresistance was measured with $B \| c$. (b) Magnetoresistance response along [$\bar{1}$100] direction at various temperatures. (c) Positive magnetoresistance with quadratic dependence on the field at 300 K. (d) Negative magnetoresistance with linear dependence on the field at 200 K.

At 250 K, in the first stage of the semiconductor regime, MR oscillates close to zero between negative and positive values. At 300 K, the MR curve is positive and the dependence on the magnetic field is quadratic. This behavior corresponds to a Weyl semimetal type I, without cone tilting.[2,48] By comparing positive MR1 and MR2 at 300 K in Fig. 3(c), one can observe that the curves are asymmetric with respect to the vertical axis. First, the minimum of each curve is subjected to a strong horizontal shift of 1.5 and 0.9 T for MR1 and MR2, respectively, and secondly, the increasing MR ratios for the left and right arms are not equal. This different contribution is attributed to the self-spin polarizer effect of the octupole clusters. The horizontal shift of the curves is larger than $H_c$ = 0.035 T of the *M-H* loop at 300 K. Thus, structural defects and ferromagnetic remanence of the canting of the Mn moments are not the origin of the shift. Furthermore, the shift is not equal for each MR$i$ indicating that the Kagome lattice does not polarize electrons at the same rate along these two directions; as the films are single crystalline, MR1 and MR2 have two spin polarized current contributions. Similar positive to negative transitions have been reported in other Weyl semimetal systems with transverse and longitudinal MR configurations.[18] In MR2 at 300 K, the left arm of the curve has a lower rate of increase that deviates from the quadratic fit. At 200 K, both the MR1 and MR2 curves, shown in Fig. 3(d), are negative and follow a linear dependence on the magnetic field. Again, there is discrepancy of the values of MR in the left and right arms of the curves for the same field, due to the anisotropic spin polarization ratios.

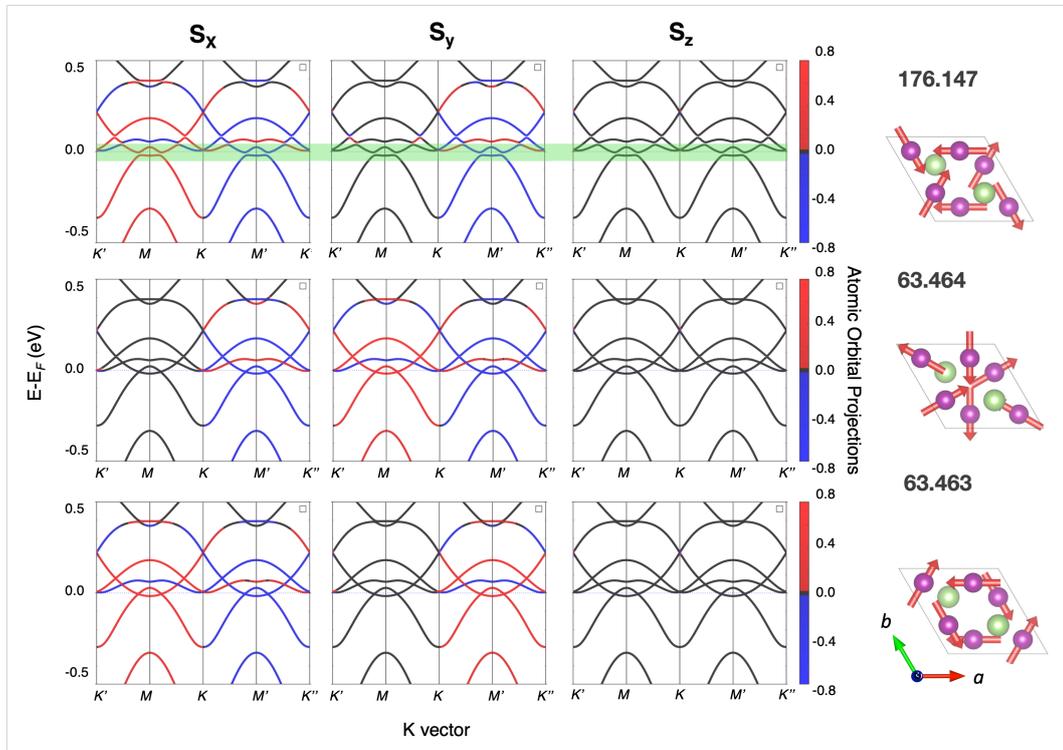

**Figure 4.** The spin-projected band structures for three spin configurations and their corresponding magnetic space groups. The blue and red represent the sign of the spin projection.

Figure 4 illustrates the correspondence to distinct magnetic space groups. For the low symmetry groups (63.463 and 63.464), we observe a metallic character near the Fermi level, although there is an opposing spin projection between $S_x$ and $S_y$. Conversely, in the high symmetry scenario with the magnetic space group 176.147, we observe a minor gap, and the spin projection Sy is opposite to that in the 63.463 group. These results agree well with the metallic to semiconductor transition observed in Fig. 2(d).[49]

In summary, $D0_{19}$-$Mn_3Ga$ epitaxial thin films were grown by magnetron sputtering on GaN (0001) substrates. RHEED and XRD patterns demostrated single crystalline quality of the films with the (0001) plane parallel to the film plane. In plane *M-H* loops lack of coercivity whereas those measured perpendicular to the film plane manifest an increase of coercivity with temperature reduction. This behavior corresponds well to the out-of-plane tilting of the Mn moments. The sample has a metallic to semiconductor transition, upon heating, at 230 K; transverse magnetoresistance was measured at different temperatures along with this transition. The magnetoresistance was found to switch from positive and quadratic with field at 300 K to negative and linear with field at 200 K, which is attributed to the change of the spin structure which in turn induces the transition from metallic to semiconductor.

## ACKNOWLEDGEMENTS

The authors acknowledge financial support from the Proyecto Interno CIMAV No. CCDPI-28/2021, from the Spanish MICIN, under PID2022-141373NB-I00 project, and from the Spanish JCCM (project CLM23-PIC-029). IMCM acknowledges to Instituto de Innovación y Competitividad ($I^2C$) from Chihuahua, Mexico, for the financial support given for this research and CONAHCYT for the support provided through the doctoral fellowship. CR acknowledges support of NSF OSI 2326754.